\documentclass[twocolumn,showpacs,unsortedaddress,showkeys,preprintnumbers,amsmath,amssymb]{revtex4}
\usepackage{graphicx}
\usepackage{dcolumn}
\usepackage{bm}

\begin{document}

\title{Model reconstruction of  nonlinear dynamical systems driven by noise}
\author{V. N. Smelyanskiy$^1$}
\email{Vadim.N.Smelyanskiy@nasa.gov}
\author{D. A. Timucin$^1$}
\author{A. Bandrivskyy$^2$}
\author{D. G. Luchinsky$^2$}

\affiliation{$^1$NASA Ames Research Center, Mail Stop 269-2,
Moffett Field, CA 94035, USA}

\affiliation{$^2$Department of Physics, Lancaster University,
Lancaster LA1 4YB, UK}

\date{\today}

\begin{abstract}
An efficient technique is introduced for   model inference of
complex nonlinear dynamical systems driven by noise. The technique
does not require extensive global optimization, provides optimal
compensation for
noise-induced errors and is robust in a broad range 
of dynamical models. It is applied to clinically measured blood
pressure signal for the simultaneous inference of the strength,
directionality,  and the noise intensities in the nonlinear
interaction between the cardiac and respiratory oscillations.
\end{abstract}


\maketitle 

Most natural and man-made systems are inherently noisy and
nonlinear. This has led  to the use of stochastic nonlinear
dynamical models  for observed  phenomena across many scientific
disciplines. Examples range from lasers \cite{laser} and molecular
motors \cite{motors},   to epidemiology \cite{epidemiology} , to
coupled matter--radiation systems in astrophysics
\cite{astrophysics}. In this approach a complex system is
characterized by projecting it onto a specific dynamical model
with  parameters obtained from the measured time-series data. In a
great number of important problems the model is not usually known
exactly from  \lq\lq first principles" and one is faced with a
rather broad range of possible parametric models to
consider. Furthermore, 
important \lq\lq hidden"  features of a model such as coupling
coefficients between the dynamical degrees of freedom  can be very
difficult to extract due to the intricate interplay between noise
and nonlinearity. These
 obstacles render the inference of stochastic nonlinear
dynamical models  from experimental time series a formidable task,
with no efficient general methods currently available for its
solution.

Deterministic inference techniques \cite{Kantz:97} consistently
fail to yield accurate parameter estimates in the presence of
noise.  The problem becomes even more complicated when both
measurement noise as well as intrinsic dynamical noise  are
present~\cite{Meyer:01}. Various numerical schemes have been
proposed recently to deal with different aspects of this inverse
problem \cite{McSharry:99,Heald:00,Meyer:00,Friedrich:00,
Meyer:01,Rossi:02,Friedrich:03}. 
 A standard approach 
is 
based on optimization of a certain cost function (a
\emph{likelihood} function) at the values of the model parameters
that best reconstruct the measurements. It can be further
generalized using a Bayesian formulation of the problem
\cite{Meyer:00,Meyer:01}.
Existing techniques usually employ numerical Monte Carlo
techniques for complex  optimization \cite{Rossi:02} or
multidimensional integration \cite{Meyer:00} tasks. Inference
results from  noisy observations are shown to be very sensitive to
the  specific choice of the likelihood function
\cite{McSharry:99}. Similarly, the correct choice of this
 function is  one of the central questions in the
inference of continuous-time noise-driven dynamical models
considered here.

In this Letter, we present an efficient technique of Bayesian
inference of  nonlinear noise-driven dynamical models from
time-series data that avoids extensive numerical optimization. It
also  guarantees optimum compensation of  noise-induced errors by
invoking the likelihood function in the form of a path integral
over the random trajectories of the  dynamical system. The
robustness of our technique in a wide range of model parameters is
verified using synthetic data from the stochastic Lorenz system.
We also present the reconstruction of a nonlinear model of
cardio-respiratory interaction from experimentally measured blood
pressure signal.

Let the trajectory ${\bf x}(t)$ of a certain $N$-dimensional
dynamical system be observed at sequential time instants $
t_0,t_1,\ldots$  and a  series ${\cal Y} = \lbrace (t_k,{\bf y}_k
);\,k=0:K\rbrace$  thus be obtained. These measurements can be
related to the (unknown)  system trajectory
through some  conditional probability distribution function (PDF)
$p_{\rm o}[{\cal Y}|{\bf x}(t)]$ giving the probability of
observing a time series ${\cal Y}$ for a specific system
trajectory ${\bf x}(t)$. If we assume that ${\bf y}_{k}$ has the
same dimension as ${\bf x}(t_k)$ and the measurement errors ${\bf
y}_k-{\bf x}(t_k)$ are uncorrelated Gaussian random variables with
mean zero and variance $\epsilon^2$, then we obtain $p_{\rm
o}\left({\cal Y}|{\cal X}\right) =\prod_{l=0}^{K}\,{\cal N}\,[{\bf
y}_{l} - {\bf
    x}(t_l),\epsilon_l]$, where ${\cal X} = \lbrace {\bf
x}(t_{k}) \rbrace$.

Assume now that the underlying dynamical system is in fact nonlinear and stochastic, evolving
according to
\begin{equation}
    \dot{\bf x}(t) = {\bf f}({\bf x}) + {\boldsymbol \xi}(t),
    \label{dynamics}
\end{equation}
\noindent where ${\boldsymbol \xi}(t)$ is an additive vector noise
process.  We parameterize this system in the following way.  The
nonlinear vector field ${\bf f}({\bf x})$ is written in the form
\begin{equation}
    {\bf f}({\bf x}) = {\hat {\bf U}}({\bf x}) \, {\bf c} \equiv {\bf f}({\bf x}; {\bf c}),
    \label{model}
\end{equation}
\noindent where ${\hat {\bf U}}({\bf x})$ is an $N \times M$
matrix of suitably chosen basis functions $\lbrace U_{n m}({\bf
x}); \,n = 1:N,\, m = 1:M \rbrace$, and ${\bf c}$ is an
$M$-dimensional coefficient vector.  An important feature of
(\ref{model}) for our subsequent development is that, while
possibly highly nonlinear in ${\bf x}$, ${\bf f}({\bf x}; {\bf
c})$ is strictly linear in ${\bf c}$.  Dynamical noise
${\boldsymbol \xi}(t)$ may also be parameterized. For instance, if
${\bf \xi}(t)$ is  stationary white and Gaussian
\begin{equation}
    \langle {\boldsymbol \xi}(t) \rangle = 0, \quad \langle {\boldsymbol \xi}(t) \, {\boldsymbol
    \xi}^{T}(t') \rangle = {\hat {\bf D}} \, \delta(t - t'),
    \label{noise}
\end{equation}
\noindent then the $N \times N$ (symmetric) noise covariance
matrix ${\hat {\bf D}}$ fully parameterizes the  noise. 
The vector elements $\lbrace c_m \rbrace$ and the matrix elements
$\lbrace D_{n n'} \rbrace$ together constitute a set ${\cal M} =
\lbrace {\bf c}, {\hat {\bf D}} \rbrace$ of unknown parameters to
be inferred from the measurements ${\cal Y}$.

In the Bayesian model inference, two distinct PDFs are ascribed to
the set of unknown model parameters: the {\em prior}
$p_{\textrm{pr}}({\cal M})$ and the {\em posterior}
$p_{\textrm{post}}({\cal M}|{\cal Y})$, respectively representing
our state of knowledge about ${\cal M}$ before and
after processing  a block of data ${\cal Y}$.  
The two PDFs are related to each other  via Bayes' theorem:
\begin{equation}
    p_{\textrm{post}}({\cal M}|{\cal Y}) = \frac{{\ell}({\cal Y}|{\cal M}) \, p_{\textrm{pr}}({\cal M})}{\int \ell({\cal Y}|{\cal
    M}) \, p_{\textrm{pr}}({\cal M}) \, {\rm d}{\cal M}}.
    \label{Bayes}
\end{equation}
\noindent Here $\ell({\cal Y}|{\cal M})$, usually termed the
\emph{likelihood}, is the conditional PDF of the measurements
${\cal Y}$ for a given choice ${\cal M}$ of the dynamical model. 
In practice, (\ref{Bayes}) can be applied iteratively using a
sequence of data blocks ${\cal Y},{\cal Y}^{\prime}$, etc. The
posterior computed from  block ${\cal Y}$ serves as the prior for
the next block ${\cal Y}^{\prime}$, etc.  For a sufficiently large
number of observations, $p_{\textrm{post}}({\cal M}|{\cal Y},{\cal
Y}^{\prime},\ldots)$ is sharply peaked 
at a certain most probable model  $\cal M={\cal M}^{\ast}$.

We  specify a prior distribution $p_{\textrm{pr}}({\cal M})$ that
is Gaussian with respect to elements of ${\bf c}$ and uniform with
respect to elements of ${\bf \hat D}$.  Thus,
$p_{\textrm{pr}}({\cal M}) = {\cal N}({\bf c}-{\bf
c}_{\textrm{pr}}, {\hat {\boldsymbol \Sigma}}_{\textrm{pr}})$,
where the mean ${\bf c}_{\textrm{pr}}$ and the covariance ${\hat
{\boldsymbol \Sigma}}_{\textrm{pr}}$ respectively encapsulate our
knowledge
 and associated uncertainty about the
coefficient vector ${\bf c}$. We now write the expression for the
likelihood
 in the form of  a path integral over the random trajectories of the system:
\begin{equation}
    \ell({\cal Y}|{\cal M}) = \int_{{\bf x}(t_{\rm i})}^{{\bf x}(t_{\rm f})} p_{\rm o}({\cal
    Y}|{\cal X}) \, {\cal F}_{\cal M}[{\bf x}(t)] \, {\cal D}{\bf x}(t),
    \label{pathint}
\end{equation}
\noindent where we choose $t_{\rm i} \ll t_{0} < t_{K} \leqslant
t_{\rm f}$ so that $\ell$ does not depend on the particular
initial and final states ${\bf x}(t_{\rm i})$, ${\bf x}(t_{\rm
f})$.  The form of the probability functional ${\cal F}_{\cal M}$
over the system trajectory ${\bf x}(t)$ is determined by the
properties of the dynamical noise ${\boldsymbol \xi}(t)$
\cite{Graham:77, Dykman:90}.

In this Letter, we are focusing  on the case of 
Gaussian white noise, as indicated in (\ref{dynamics}),
(\ref{noise}). We  consider a uniform sampling scheme $t_{k} =
t_{0} + h k$, $h \equiv (t_{K} - t_{0})/K$ and assume  that for
each trajectory component $x_n(t)$ the measurement error
$\epsilon$ is negligible compared with the fluctuations induced by
the dynamical noise; that is, $\epsilon^2\ll h ({\bf \hat
D}^2)_{n\,n}$. Consequently, we  use $p_{\rm o}({\cal Y}|{\cal X})
\rightarrow \prod_{k = 0}^{K} \delta[{\bf y}_{k} - {\bf x}(t_k)]$
in (\ref{pathint}).   Using results from \cite{Graham:77} for
${\cal F}_{\cal M}[{\bf x}(t)]$, the logarithm of the likelihood
(\ref{pathint}) takes the following form for sufficiently large
$K$ (small time step $h$):
\begin{eqnarray}
 && \hspace{-0.3in}-\frac{2}{K}\log  \ell({\cal Y}|{\cal M})  =
\ln\det{\hat{\bf D}}
+\frac{h}{K}\sum_{k=0}^{K-1}\left[\,\,\mathop{\rm
    tr}{\hat {\bf \Phi}}({\bf y}_{k}; {\bf c})\right.\label{likelihood} \\
  &&\hspace{-0.2in}\left. +(\dot{\bf y}_{k}  - {\bf f}({\bf y}_{k}; {\bf c}))^T \, {\hat {\bf D}}^{-1} \,
    (\dot{\bf y}_{k}  - {\bf f}({\bf y}_{k}; {\bf c}))\right]+ N\ln(2\pi
    h)
    ,\nonumber
\end{eqnarray}
\noindent \noindent here we introduce the \lq\lq velocity" $\dot
{\bf y}_{k}$ and matrix ${\bf \hat \Phi}({\bf x})$
\[\dot {\bf y}_{k}\equiv h^{-1} ({\bf y}_{k+1}-{\bf y}_k),\quad
({\hat {\bf \Phi}}({\bf x}; {\bf c}))_{n\,n'}\equiv
\partial f_{n}({\bf x}; {\bf c})/\partial x_{n'}.\]\noindent
With the use of (\ref{model}), substitution of the  prior
$p_{\textrm{pr}}({\cal M})$ and the  likelihood $\ell({\cal
Y}|{\cal M})$
 into (\ref{Bayes}) yields the posterior
$p_{\textrm{post}}({\cal M}|{\cal Y}) ={\rm const}\times
\exp[-S({\cal M}|{\cal Y})]$, where

\begin{equation}
    S({\cal M}|{\cal Y})\equiv S_{\textsf{y}}({\bf c},{\bf \hat D}) =
    \frac{1}{2}\rho_{\textsf{y}}({\bf \hat  D}) - {\bf c}^{T} {\bf w}_{\textsf{y}}({\bf \hat  D}) +
     \frac{1}{2}
    {\bf c}^{T} {\bf \hat  \Xi}_{\textsf{y}}({\bf \hat  D}) {\bf c}.
    \label{action}
\end{equation}

\noindent Here,  use was made of the definitions
\begin{eqnarray}
 &&\hspace{0.045in}\rho_\textsf{y}({\bf \hat{D}}) =  h \, \sum_{k = 0}^{K - 1} \dot{{\bf
 y}}_{k}^{T}
    \, {\bf \hat D}^{-1} \, \dot{{\bf
 y}}_{k}  + K \, \ln (\det
    { \bf \hat D}),  \label{defs} \\ &&{\bf w}_\textsf{y}({\bf \hat D})  =
    {\bf \hat \Sigma}_{\textrm{pr}}^{-1} \, {\bf c}_{\textrm{pr}}
    + h\sum_{k = 0}^{K - 1}\left[ {\bf \hat U}_{k}^T \, {\bf \hat D}^{-1} \, \dot{{\bf
 y}}_{k}-  \frac{1}{2}  {\bf v}({\bf y}_k)\right],
   \nonumber \\ &&{\bf \hat  \Xi}_\textsf{y}({\bf \hat  D})
    =  {\bf \hat  \Sigma}^{-1}_{\textrm{pr}} + h \, \sum_{k = 0}^{K - 1} {\bf \hat U}_{k}^{T} \, {\bf \hat D}^{-1} \,
    {\bf \hat U}_{k},
    \nonumber
\end{eqnarray}
\noindent where $ { \bf \hat U}_{k} \equiv {\bf \hat U}({\bf
y}_{k})$ and  the components of  vector ${\bf v}({\bf x})$ are:
\begin{equation}
\textrm{v}_{m}({\bf x})=\sum_{n=1}^{N}\frac{\partial U_{n\,m}({\bf
x})}{\partial x_n},\quad m=1:M.\label{v}
\end{equation}
\noindent

The mean values of ${\bf c}$ and ${\bf \hat D}$ in the posterior
distribution give the best estimates for the model parameters for
a given block of
data ${\cal Y}$ of  length $K$ and provide a global minimum to 
$ S_\textsf{y}({\bf c}, {\hat {\bf D}})$. We  handle this
optimization problem in the following way. Assume for the moment
that ${\bf c}$ is known in (\ref{action}). Then the posterior
distribution over ${\bf \hat D}$ has a mean  ${\bf\hat D}^{\bf
\prime}_{\textrm{post}}={\bf \hat \Theta}_{\textsf{y}}({\bf c})$
that provides a  minimum to $S_\textsf{y}({\bf c},{\bf \hat D})$
with respect to ${\bf \hat D}={\bf \hat D}^T$.   Its matrix
elements are
\begin{equation}
\hspace{-0.01in}{\bf \hat \Theta}_{\textsf{y}}^{n n'}({\bf c})
\equiv \frac{1}{K} \, \sum_{k=0}^{K-1} \left[ {\dot {\bf y}}_{k} -
    {\hat {\bf U}}({\bf y}_{k}) \, {\bf c} \right]_n \left[ {\dot {\bf y}}_{k} - {\hat {\bf U}}({\bf
    y}_{k}) \, {\bf c} \right]^{T}_{n'}.\label{updateD}
\end{equation}
\noindent Alternatively, assume next that
 ${\hat {\bf D}}$ is known, and note
from (\ref{action}) that in this case the posterior distribution
over ${\bf c}$ is Gaussian. Its covariance is given by ${\bf
\hat\Xi}_\textsf{y}({\bf \hat D})$ and the mean ${\bf
c}^{\prime}_{\textrm{post}}$ minimizes  $S_\textsf{y}({\bf c},{\bf
\hat D})$ with respect to ${\bf c}$
\begin{equation}
{\bf c}^{\prime}_{\textrm{post}}={\hat {\boldsymbol
\Xi}}^{-1}_\textsf{y}({\bf \hat D}){\bf w}_\textsf{y}({\bf \hat
D}).\label{updateC}
\end{equation}
\noindent 
We repeat  this two-step optimization procedure iteratively,
starting from some prior values ${\bf c}_{\textrm{pr}}$ and ${\bf
\hat \Sigma}_{\textrm{pr}}$. We do not need a prior for ${\bf D}$
at this stage, according to (\ref{defs})-(\ref{updateC}). At
convergence we obtain the \lq\lq true" mean posterior values,
${\bf c}^{\prime}_{\textrm{post}}\rightarrow {\bf
c}_{\textrm{\textrm{\textrm{post}}}}$  and ${\bf \hat D}^{\bf
\prime}_{post}\rightarrow {\bf \hat D}_{\textrm{post}}$. The
posterior covariance matrix ${\bf \hat
\Sigma}_{\textrm{post}}={\bf \hat \Xi}_{y}^{-1}({\bf \hat
D}_{\textrm{post}}) $.

 To continue the inference process with a
new block of data ${\cal Y}^{\prime}$ of  length $K^{\prime}$ we
update the prior mean ${\bf c}_{\textrm{pr}}= {\bf
c}_{\textrm{post}}$, ${\bf \hat D}_{\textrm{pr}}={\bf \hat
D}_{\textrm{post}}$,  and covariance, ${\hat {\boldsymbol
\Sigma}}_{\textrm{pr}} = {\hat {\boldsymbol
\Sigma}}_{\textrm{post}}$, and repeat the two-step optimization
procedure. The  modification is that updates    now explicitly
involve  ${\bf \hat D}_{\textrm{pr}}$
\begin{equation}
{\bf \hat D}_{\textrm{post}} =\frac{K}{K^{\prime}+K}{\bf \hat
D}_{\textrm{pr}}+\frac{K^{\prime}}{K^{\prime}+K}{\bf \hat
\Theta}_{\,y'}({\bf c}_{\textrm{post}}).\label{updateD1}
\end{equation}
We obtain (\ref{updateD1}) from (\ref{updateD}) where the data
record ${\cal Y}\cup{\cal Y}'$ of  length $K+K'$ is used instead
of ${\cal Y}$ and the sum over the first $K$ data points (block
${\cal Y}$) is given by the ${\bf \hat D}_{\textrm{pr}}$. Clearly,
many non-overlapping, and not necessarily contiguous, data blocks
of varying lengths  may be used in this recursive model inference
algorithm \cite{note}.

The  terms involving ${\rm tr}{\bf \hat \Phi}({\bf y}_k)$ in
(\ref{likelihood}) originate from the prefactor in 
${\cal F}_{\cal M}[{\bf x}(t)]$ (\ref{pathint}) and do not vanish
at the dynamical system attractors (\ref{dynamics}), unlike the
terms in (\ref{likelihood}) involving $ \dot{\bf y}_k-{\bf f}({\bf
y}_k;{\bf c})$. Therefore both types of terms are required  to
optimally balance the effect of noise effect in $\{{\bf y}_k \}$
(\ref{action}) and provide the robust convergence.

\begin{figure}[ht!]
\includegraphics[width=7.2cm,height=5cm]{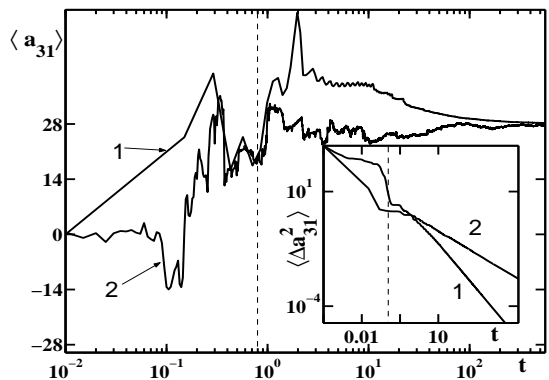}
\caption{\label{fig:lorenz} Examples of convergence of the
coefficient $a_{31}$ corresponding to parameter $r$ of the Lorenz
system (\ref{Loretnz}) for the total length of the time record
$T=560$ and the following sets of parameters: line $1$
$\{d_n\}=\{0.01, 0.012, 0.014\}$ time step $h=0.002$; line $2$
$\{d_n\}=\{100, 120, 140\}$ and $h=0.00002$. The insert shows
dispersion $\langle\Delta a_{31}^2\rangle$=$\langle
a_{31}^2\rangle$ - $\langle a_{31}\rangle^2$ for the same sets of
parameters. The vertical
 dashed line shows the time-scale of the step-wise
decrease in the variance.}
\end{figure}

We now consider an example of the dynamics in (\ref{dynamics})
given by a noise-driven chaotic Lorenz system. It has dynamical
variables, ${\bf x}=\{x_1,x_2,x_3\}$, the
 vector field
\begin{equation} {\bf
f}({\bf x})= \left(x_{2} - x_1,\,\,\,r  x_1 - x_2 - x_1
x_{3},\,\,\,x_1 x_2 - b x_3\right),\label{Loretnz}
\end{equation}
\noindent and  noise correlation matrix
$\langle\xi_{n}(t)\xi_{n'}(t')\rangle=d_n\delta_{n,n'}$. The
parameters in (\ref{Loretnz}) are $\sigma = 10$, $r = 28$, $b =
\frac{8}{3}$. We sample a system trajectory ${\bf x}(t)$ and
produce a \lq\lq data record" $\{{\bf y}(t_k)\}$  to be fed
directly into the algorithm.  As an inferential framework, we
introduce the following (bilinear) model of stochastic dynamics
for ${\bf x}(t)$:
\begin{equation}
    {\dot x}_{n} = \sum_{i = 1}^{3} a_{n i} \, x_{i}(t) + \sum_{i < j =
    1}^{3} b_{n i j} \,
    x_i(t) \, x_j(t) + \xi_{n}(t),
    \label{mod:lorenz}
\end{equation}
\noindent $n, i, j = 1, 2, 3$, where ${\cal M}=\{\{a_{n i}\},
\{b_{n i j}\}, \{D_{n n'}\}\}$ is the vector of 18 unknown
coefficients. The form of the 12 basis functions $U_{n m}({\bf
x})$ is evident from (\ref{mod:lorenz}).  We were able to infer
the accurate values of ${\bf c},{\bf \hat D}$ for time step $h$
varying from $0.01$ to $10^{-6}$ and noise intensities $d_n$
varying from 0 to $10^{2}$. An example of convergence of the
coefficients is shown in  Fig. \ref{fig:lorenz}.
We found a step-wise decrease in variances that occurs on a time
scale of the period of oscillations $\tau_{osc}\approx 0.6$
(dashed line in the figure).  The error of the inference is
sensitive to the noise intensity, total time $T$ and the time step
$h$. For example, for the parameters of the  curve \emph{1} in the
Fig. \ref{fig:lorenz} the relative error was $0.015\%$. The ratio
$T/h$ has to be increased at least 250 times to achieve error less
then $1\%$ when the noise intensity is increased $10^4$ times
(curve \emph{2} in the figure).

\begin{figure}
\includegraphics[width=7.2cm,height=6cm]{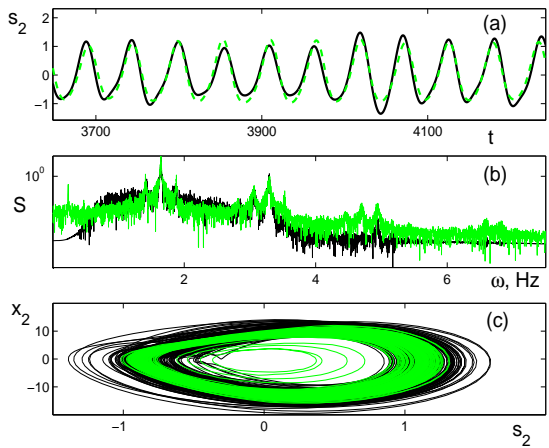}
\caption{\label{fig:cardiac} (a) Time series of the cardiac
oscillations $s_2(n)$ in arbitrary units (black line) obtained
from central venous blood pressure with 4-th order Butterworth
filter (low-and high- cut-off frequencies were $f_l=0.8Hz$ and
$f_h=2.8Hz$ and the sample rate was 90 Hz after resampling of the
original signal). Inferred time series of the cardiac oscillator
(green line). (b) Power spectrum of the cardiac oscillations
obtained from the real data (black line). Power spectrum of the
inferred oscillations (green line). (c) Limit cycle of the cardiac
oscillations $(x_2(n),y_2(n)$ obtained from real data as described
in the text (black line). Limit cycle of inferred oscillations
(green line).}
\end{figure}

Finally, we apply our method to  study the  stochastic nonlinear
dynamics of complex physiological system. To be specific we infer
the strength, directionality and a degree of randomness of the
cardiorespiratory interaction from the central venous blood
pressure signal (record 24 of the MGH/MF Waveform Database
available at www.physionet.org). Such estimations provide valuable
diagnostic information about the responses of the autonomous
nervous system \cite{Hayano:03,Malpas:02}. However, it is
inherently difficult to dissociate a specific response from the
rest of the cardiovascular interactions and the mechanical
properties of the cardiovascular system in the intact
organism~\cite{Jordan:95}. Therefore a number of numerical
techniques were introduced to address this problem using e.g.
linear approximations~\cite{Taylor:01}, or semi-quantitative
estimations of either the strength of some of the nonlinear
terms~\cite{Jamsek:03} or the directionality of
coupling~\cite{Rosenblum:02,Palus:01}. But the problem remains
 wide open because of the complexity and nonlinearity of the
cardiovascular interactions. Our algorithm provides an alternative
effective approach to the solution of this problem. To demonstrate
this we use a combination of low- and high-pass Butterworth
filters to decompose the blood pressure signal into 2-dimensional
time series $\lbrace {\bf s}(t_k)=(s_0(t_k),s_1(t_k)),\,\,t_k=k
h,\,k=0:K\rbrace$ representing  observations of mechanical cardiac
and respiratory degrees of freedom on a discrete time grid  with
  step $h=0.002$ sec. 
We now
introduce an auxiliary two-dimensional dynamical system whos
trajectory ${\bf x}(t)=(x_0(t),x_1(t))$ is related to the
observations $\{{\bf s}(t_k)\}$  as follows
\[x_n(t_k) = a_{1n}\,\frac{s_n(t_k+h)-s_n(t_k)}{h} +a_{2n} s_n(t_k)+a_{3n},\]
where $n=0,1$. The corresponding simplified model of the nonlinear
interaction between the cardiac and respiratory limit cycles has
the form (cf. with ~\cite{Stefanovska:01a})
\begin{eqnarray}
 &&\hspace{-0.15in}\dot x_n =
 b_{1n}+b_{2n}s_n+b_{3n}x_n+b_{4n}s_n^2+b_{5n}x_n^2
 + b_{6n}s_n x_n \nonumber \\
 &&\hspace{-0.15in}+ \,b_{7n}s_n^3+b_{8n}s_n^2x_n+b_{9n}x_n
 x_n^2+b_{10n}x_n^3+ b_{11n}x_n x_{n-1} \nonumber
 \\&&\hspace{-0.15in}+\, b_{12n} x_n^2 x_{1-n}+ b_{13n} x_n x_{1-n}^{2} +\xi_{n}(t),
 \quad n=0,1.\label{eq:cardiorespiratory}
\end{eqnarray}
\noindent where $\xi_n(t)$ is a  Gaussian white noise with
correlation matrix $D_{n\,n'}$ (\ref{noise}). We emphasize that a
number of important parameters of the decomposition of the
original signal (including the bandwidth, the order of the filters
and the scaling parameters $a_{ki}$) have to be selected to
minimize the cost (\ref{action}) and provide the best fit to the
measured time series $\{ {\bf s}(t_k)\}$. The parameters of the
model (\ref{eq:cardiorespiratory}) can now be inferred directly
from the noninvasively measured time series of blood pressure. The
comparison between the time series of the inferred and actual
cardiac oscillations is shown in  Fig. \ref{fig:cardiac}.
Similar results are obtained for the respiratory oscillations. In
particular, the parameters of the nonlinear coupling and of the
noise intensity  of the cardiac oscillations are $b_{112}=3.9,
b_{122}=0.62$, $b_{132}=-13.4$, and $D_{22}=4.75$
($\langle\xi^2_{y_2}(t)\rangle=D_{22}$). Consistent with
expectations, in all experiments the  parameters of the nonlinear
coupling are two orders of magnitude higher for the cardiac
oscillations as compared to their values for the respiratory
oscillations 
reflecting the fact that respiration strongly modulates cardiac
oscillations, while the opposite effect of the cardiac
oscillations on  respiration is weak. Remarkably, our technique
infers simultaneously the strength, directionality of coupling and
the noise intensities in the cardiorespiratory interaction
directly from the non-invasively measured time series.

In conclusion, we have derived an efficient technique for
recursive  inference of dynamical model parameters that does not
require extensive numerical optimization and provides optimum
compensation for the dynamical noise-induced errors. We
verified the robustness of the  technique in a very
broad range of parameters of dynamical models, using synthetic
data from the chaotic noise-driven Lorenz system. Successful
application of the technique to  inference from real data of the
nonlinear interaction between the cardiac and respiratory
oscillations in the human cardiovascular system is particularly
encouraging, as it opens up a new avenue for the Bayesian
inference of strongly nonlinear and noisy dynamical systems with
limited first-principles knowledge. Future extensions will include   
more realistic observation schemes with \lq\lq hidden" variables
and finite measurement noise.

This work was supported by NASA  IS IDU project (USA) and by EPSRC
(UK).

\end{document}